\documentclass[a4paper,fleqn,usenatbib]{mnras}

\usepackage{newtxtext,newtxmath}

\usepackage[T1]{fontenc}
\usepackage{ae,aecompl}

\usepackage[dvipsnames]{xcolor}
\usepackage{float}
\usepackage{ulem}

\usepackage{graphicx}	
\usepackage{amsmath}	
\usepackage{amssymb}	

\defcitealias{lee2012}{Paper~1}
\defcitealias{lee2016}{Paper~2}

\title[Optical spectroscopy of young tidal objects]{Optical spectroscopy of young tidal objects around two interacting galaxy pairs}

\author[K. Lee-Waddell et al.]{
K. Lee-Waddell$^{1}$\thanks{E-mail: karen.lee-waddell@csiro.au},
J.P. Madrid$^{1}$,
K. Spekkens$^{2}$,
C.J. Donzelli$^{3}$,
B.S. Koribalski$^{1}$,
\newauthor 
P. Serra$^{4}$,
J. Cannon$^{5}$
\\
$^{1}$CSIRO Astronomy and Space Science, Australia Telescope National Facility, PO Box 76, Epping, NSW 1710, Australia\\
$^{2}$Department of Physics, Royal Military College of Canada, PO Box 17000, Station Forces, Kingston, ON K7K 7B4, Canada\\
$^{3}$Instituto de Astronom\'ia Te\'orica y Experimental, CONICET-UNC, Laprida 922, C\'ordoba, Argentina; Observatorio Astron\'omico de C\'ordoba, \\
\hspace{5mm}UNC, Laprida 854, C\'ordoba, Argentina\\
$^{4}$INAF -- Osservatorio Astronomico di Cagliari, Via della Scienza 5, I-09047 Selargius (CA), Italy\\
$^{5}$Department of Physics and Astronomy, Macalester College, 1600 Grand Avenue, Saint Paul, MN 55105, USA\\
}

\date{Accepted 2018 July 26. Received 2018 July 26; in original form 2018 January 23}

\pubyear{2018}

\begin{document}
\label{firstpage}
\pagerange{\pageref{firstpage}--\pageref{lastpage}}
\maketitle

\begin{abstract}
We present Gemini Multi-Object Spectrograph (GMOS) spectra of {H\sc{i}}-rich tidal features located around the outskirts of two interacting galaxy pairs, NGC~3166/9 and NGC~4725/47.  These follow-up observations are part of a multi-wavelength campaign to study the properties and frequency of tidal dwarf galaxies (TDGs) in group environments.  Based on the calculated gas-phase metallicity and redshift, in addition to the previously measured {H\sc{i}} and stellar properties, we have confirmed the tidal origins of TDG candidate AGC~208457, which has emerged from the tidal debris of an interaction between the NGC~3166/9 galaxy pair.  By comparing {H\sc{i}} and optical recessional velocities, we have also confirmed a physical association of the gaseous knots and star clusters embedded in the metal-rich tidal tail of NGC~4747.
\end{abstract}

\begin{keywords}
galaxies: dwarf -- galaxies: groups: individuals: NGC 3166/9 -- galaxies: groups: individual: NGC 4725/47 -- galaxies: interactions
\end{keywords}



\section{Introduction}

Tidal interactions dominate the dynamical processing of galaxies in low- to intermediate-density environments \citep{bla2009}.  Interactions between gas-rich galaxies can produce prominent tidal features such as transitory tails and knots as well as long-lived tidal dwarf galaxies (TDGs; \citealt{bou2006}).  These features can greatly constrain the interaction history and properties of the host galaxies and enable in depth study of this evolutionary process (e.g.~\citealt{too1972}; \citealt{bou2007}).  

Numerical simulations suggests that TDGs are self-gravitating galaxies that form from the outer disk material of larger galaxies.  Accordingly, they can inherit the chemical abundances of the parent galaxies and are typically more metal-rich than classical dwarfs \citep{lis2016}. Various studies using optical spectroscopy, find that stellar clumps located within tidal tails and TDGs generally have oxygen abundance metallicities of 12+log(O/H)~>~8.3 and some can be as metal-rich as 12+log(O/H)~=~8.8 (\citealt{duc1998}, \citealt{dem2012}, \citealt{duc2014}, \citealt{lel2015}).  Whereas, typical dwarf irregular galaxies have metallicities of 7.3~<~12+log(O/H)~<~8.4 (\citealt{van2006}, \citealt{cro2009}, \citealt{ber2012}).

While classical dwarf galaxies are generally considered to be dark matter dominated, TDGs are predicted to have little to no associated dark matter \citep{bou2006}.  The latter is based on predictions using Lambda cold dark matter ($\Lambda$CDM) cosmology, as dark matter is mostly located in outer halo of galaxies rather than in the gaseous disk region where TDGs are formed. However, it is difficult to establish if TDG candidates are actually dark matter free as total mass measurements generally assume dynamical equilibrium and showing that these objects are virialized is quite challenging (see \citealt{lel2015} and \citealt{flo2016} for further details).  Only a handful of TDGs have been detected and spectroscopically confirmed as authentic (e.g.~\citealt{duc2014}; \citealt{lel2015}; \citealt{lis2016}), which prevents detailed analysis of this class of objects on a universal scale.  As such, the significance of TDGs and whether they can be used to improve the standard model of cosmology remains a topic of debate (e.g.~\citealt{kro2012}, \citealt{flo2016}).  

We are carrying out a multi-wavelength investigation of nearby galaxy groups to find TDG candidates and study their properties.  Specifically, we used high-resolution neutral hydrogen ({H\sc{i}}) observations from the Giant Metrewave Radio Telescope (GMRT) to identify gas-rich dwarfs and measure their dynamical properties (\citealt{lee2012}; hereafter, \citetalias{lee2012}).  Optical imaging, from the Canada-France-Hawaii Telescope (CFHT) MegaCam, detected putative low-surface brightness stellar counterparts of the {H\sc{i}} tidal features and enabled stellar mass measurements (\citealt{lee2016}; hereafter, \citetalias{lee2016}).  The combination of {H\sc{i}} mapping and optical photometry provided dynamical-to-baryonic mass estimates for the initial classification of tidal objects.  

\begin{figure*}
\begin{center}
	\includegraphics[width=170mm]{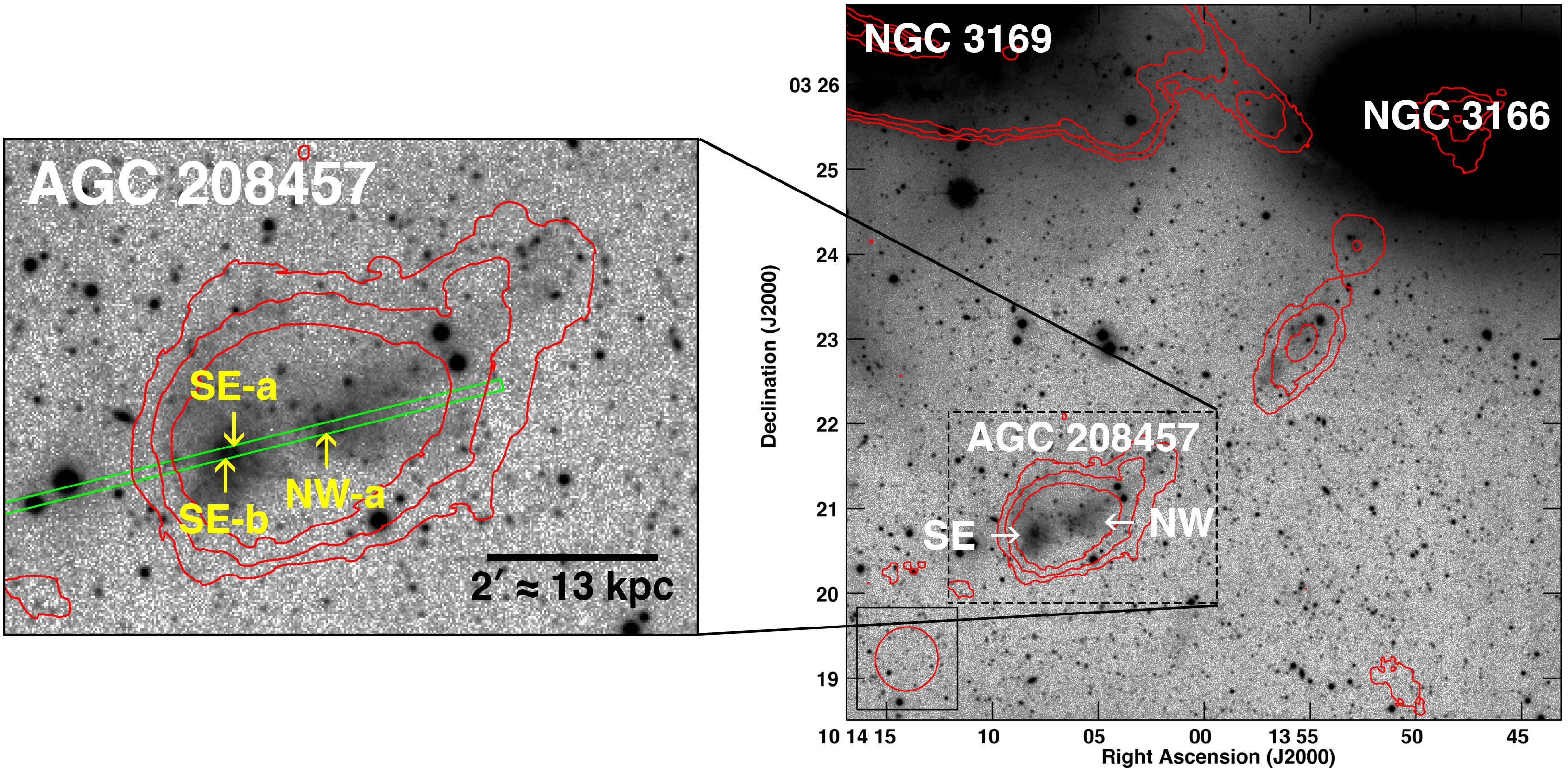}
	\caption{GMRT total H{\sc{i}} intensity contours at $N_{H_I} = (0.9, 1.5, 2.3) \times 10^{20}$ atoms cm$^{-2}$, in red, superimposed on a CFHT MegaCam \textit{g'}-band image of the central region between NGC~3166 and NGC~3169 (reproduced from \citetalias{lee2012} and \citetalias{lee2016}).  The two stellar clumps of AGC~208457 are indicated.  The 45 arcsec synthesized GMRT beam is shown in the bottom-left corner.  Zooming in on AGC~208457, the green box shows the location of the GMOS-N slit (illustrated with twice the actual width) and the yellow arrows indicate the approximate locations of the extracted spectra.  The angular size scale assumes a distance of 23.2 Mpc to the NGC 3166/9 group \citep{tul2013}.
\label{fig:TDG}}
\end{center}
\end{figure*}

The most promising TDG candidate identified by our campaign thus far is AGC~208457, located around the NGC~3166/9 galaxy pair.  Our GMRT {H\sc{i}} measurements indicate that AGC~208457 has sufficient mass ($M_{H_I} = 2.3 \pm 0.3 \times 10^8$ M$_{\odot}$; \citetalias{lee2012}) to become a long-lived TDG \citep{bou2006}.  Assuming that this object is self-gravitating and virialized, it has a dynamical-to-gas mass ratio close to unity denoting a lack of dark matter \citepalias{lee2012}.  Deep optical photometry in \citetalias{lee2016} shows two stellar clumps with a combined mass of $M_{\mbox{stellar}} \sim 2 \times 10^7$ M$_{\odot}$ spatially located within the highest {H\sc{i}} column density region of AGC~208457 (see Figure~\ref{fig:TDG}).  These stars appear to be fairly blue and are bright in the ultraviolet (UV) suggesting that they are quite young (\citetalias{lee2012}, \citetalias{lee2016}).  AGC~208457 therefore has many hallmarks of a TDG.  Confirming a high metallicity for this object would add further evidence for its tidal origin.

\begin{figure*}
\begin{center}
	\includegraphics[width=170mm]{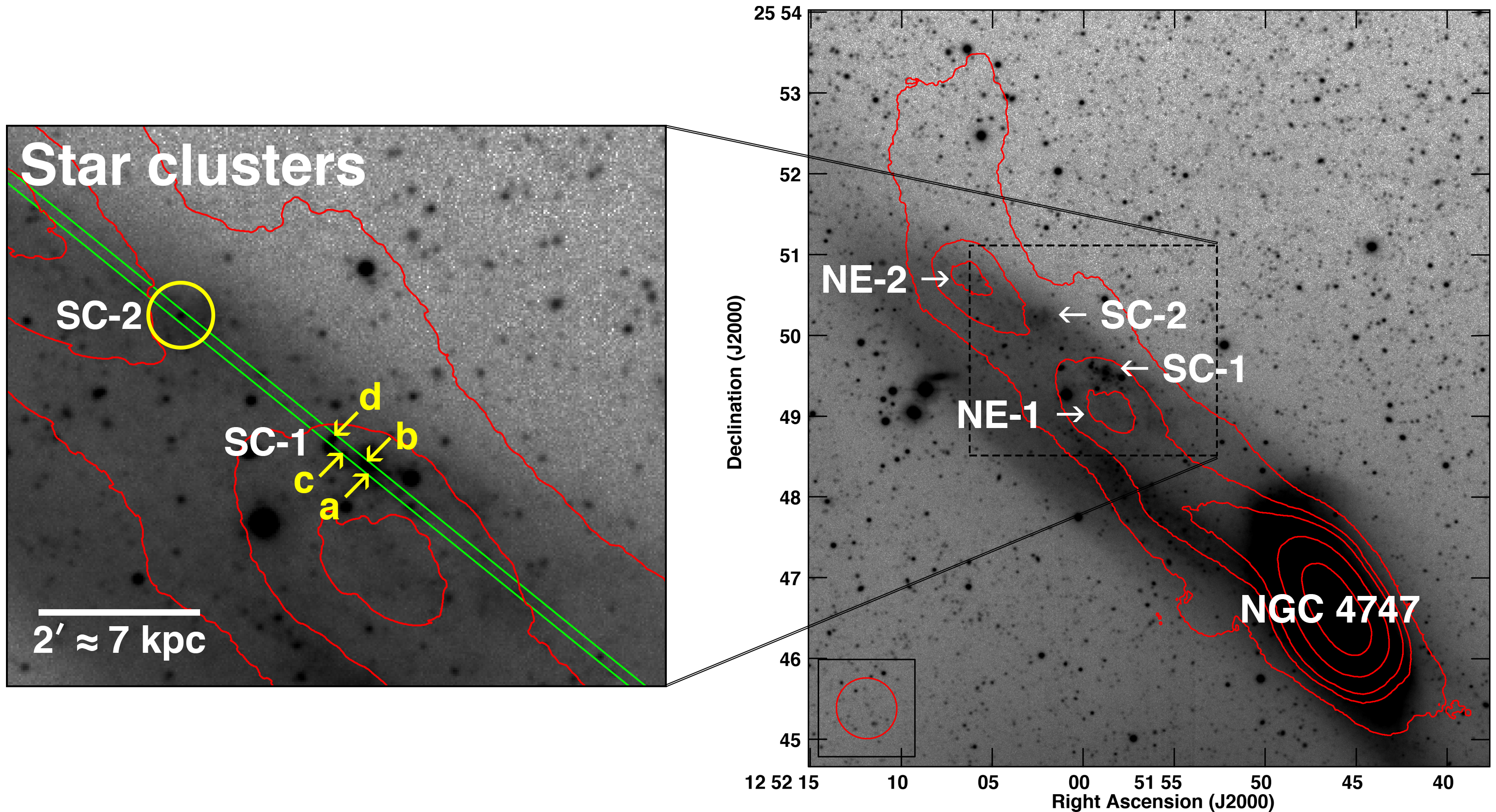}
	\caption{GMRT total H{\sc{i}} intensity contours at $N_{H_I} = (2, 5, 7, 12, 20) \times 10^{20}$ atoms cm$^{-2}$, in red, superimposed on a CFHT MegaCam \textit{g'}-band image of NGC 4747 and the features in its gas-rich tail (reproduced from \citetalias{lee2016}).  The 45 arcsec synthesized GMRT beam is shown in the bottom-left corner.  Zooming in on star clusters, the green box shows the location of the GMOS-N slit (illustrated with twice the actual width) and the yellow arrows indicate the approximate locations of the extracted spectra from star cluster SC-1.  Star cluster SC-2 (denoted by the yellow circle) was too faint for further analysis.  The angular size scale assumes a distance of 12.6 Mpc to the NGC 4725/47 group \citep{tul2013}.
\label{fig:knots}}
\end{center}
\end{figure*}

Two other interesting TDG candidates are found in the eastern tidal tail of NGC~4747 \citepalias{lee2016}.  Gravitational interactions between neighbouring spiral galaxies, NGC~4725 and NGC~4747, likely resulted in gas-rich tidal tails.  The most prominent tail extends about 8 arcmin ($\sim$30 kpc) to the north-east of NGC~4747 and appears to contain two distinguishable star clusters (SCs) with masses of $M_{\mbox{stellar}} \sim 6 \times 10^6$ and $9 \times 10^6$ M$_{\odot}$ \citepalias{lee2016}.  As shown in Figure~\ref{fig:knots}, the {H\sc{i}} tail is oriented at a slight angle from a bent stellar tail and contains two {H\sc{i}} peaks (referred to as NE-1 and NE-2) each with a mass of M$_{H_I}\approx10^8$ $M_{\odot}$ \citepalias{lee2016}.  One of the SCs (i.e.~SC-1) approximately coincides spatially with NE-1.  Galaxy Evolution Explorer ($GALEX$) UV data, from the GR6/7 public data release, reveal a faint arc of recent star-formation that spatially aligns with the two SCs but that is also offset from the {H\sc{i}} \citepalias{lee2016}.  If the {H\sc{i}} and stellar features have similar recessional velocities, which would imply that they are located at similar distances, then these features would be indicative of the early stages of TDG formation and/or turbulence resulting from a recent interaction event (see \citealt{may2007}; \citealt{sen2015}).

To confirm a physical association between the {H\sc{i}} detections and the stellar components (i.e.~verify the distance to each source in order to rule out foreground/background contaminants) of AGC~208457 and the SCs in the tail of NGC~4747 as well as measure metallicities to authenticate tidal origins, we obtained long-slit optical spectroscopy using the Gemini Multi-Object Spectrograph (GMOS).  Our spectroscopic observations are described in Section \ref{sec:obs}.  In Section \ref{sec:results}, we employ the same method as \cite{duc2014}, who confirm a TDG, NGC~5557-E1, with an oxygen abundance of 12+log(O/H)~=~8.6, to measure metallicities.  We discuss our findings in Section \ref{sec:discuss} and present our final conclusions in Section \ref{sec:conclude}.

\section{New Gemini Observations}
\label{sec:obs}

We obtained new spectra of AGC 208457 and the gas-rich tail of NGC~4747 with GMOS on Gemini North. GMOS-North is an {\sc e2v} CCD detector array built with three CCD chips that produces files with six extensions.  The observations were obtained using GMOS longslit mode with a 1.5$\arcsec$-wide slit and the B600\_G5307 grating of R = 1700, which has 600 lines per mm ruling density and a blaze wavelength of 461 nm, centered to cover the H$\alpha$, H$\beta$, [{N\sc{ii}}] and [{O\sc{iii}}] emission lines.  We use a CCD binning of 2 $\times$ 2 pixels resulting in 0.90\AA~spectral resolution and a spatial resolution of 0.145 arcseconds.

For AGC 208457, the slit was positioned along its major axis, across the two stellar clumps that are clearly visible in the optical images (see Figure~\ref{fig:TDG}). These features coincide with the highest {H\sc{i}} column density region of AGC~208457 and appear to be the most active sites of star formation in this TDG candidate.  For the tail of NGC~4747, the slit was aligned with the two SCs shown in Figure~\ref{fig:knots}.  SC-1 is located within the {H\sc{i}} knot, NE-1, albeit slightly offset from the centre of the gas peak.  

The observations were taken over three nights.  Half of the integration time for each science source was taken at a central wavelength of 5550\AA, while other half was at 5600\AA.  Additionally, an offset of 15 arcseconds between science exposures was implemented.  These minor shifts mitigate the presence of CCD gaps in the final spectra.  Sky conditions and other observing details are summarized in Table~\ref{table:obs}.  The stellar feature of AGC~208457 has an apparent $r'$-band magnitude of $m_{r}=18.5$ mag \citepalias{lee2016} and required significantly more integration time than the SCs within the tail of NGC~4747 ($m_{r}$ = 17.1 mag and $m_{r}$ = 17.6 mag for SC-1 and SC-2 respectively) to achieve the estimated signal-to-noise for the weakest emission line (i.e.~H$\beta$; see Section~\ref{sec:results}) required to compute metallicities.  On 24 November, sky conditions worsened during the fourth exposure.  As such, an additional exposure of AGC~208457 was taken in December.  For calibration purposes, observations of the spectrophotometric standard star, EG131, which enable extinction correction, and arcs with the copper-argon CuAr lamp were also carried out during this observing program.

\begin{table*}
 \centering
 \begin{minipage}{110mm}
 \caption{Details of GMOS-N observations}
 \label{table:obs}
\begin{tabular}{ l c c c}  
\hline									
Date									&24 Nov 2016		&29 Dec 2016		&9 Jan 2017\\
\hline									
Science target							&AGC 208457		&AGC 208457		&NGC~4747 tail\\
Atmosphere stability / image quality (IQ \%)	&70				&70				&70\\
Cloud coverage (CC \%)					&70				&70				&50\\
Water vapour / transparency (WV \%)		&80				&50				&80\\
Sky darkness / backgound (SB \%)			&50				&20				&20\\ 
Number of exposures					&4				&1				&4\\
Target exposure time (s)		&1735			&1735			&500\\
Total integration time (s)					&6940			&1735			&2000\\
\hline																	
\end{tabular} 
\end{minipage}
\end{table*}

\subsection{Data Processing}

All science and calibration data were retrieved from the Gemini archive and reduced using the standard procedure based on the successive application of the following IRAF tasks within the Gemini package. This data reduction method follows the procedure presented in \citet{mad2013}. Flatfields were processed with {\sc gsflat} to create a flat that was then applied to the science data.  Running the task {\sc gsreduce} enabled bias and overscan subtraction as well as cosmic ray removal. The task {\sc gsreduce} was also used to flatfield the data and the task {\sc gmosaic} combined the different file extensions created by detector architecture into a single frame.  Sky subtraction was carried out with {\sc gsskysub} by defining a region of the CCD frame free of flux from the science spectra to derive the sky level. 

Spectra from the CuAr arcs were processed with {\sc gswavelength} in order to build the wavelength calibration. The task {\sc gstransform} was then used to rectify, calibrate, and interpolate the spectra following the solutions derived with {\sc gswavelength}. The residual calibration error after wavelength correction using the CuAr arcs is $\sim$0.3\AA.  Spectra of the science targets were extracted using the task {\sc gsextract} without applying any weights (i.e.~using a simple sum of pixel values along the spectrum). The width of the extraction in the spatial direction is equivalent to the seeing of the image. Flux calibration and extinction correction of those spectra were carried out using the spectra of EG131 and the tasks {\sc gsstandard} and {\sc gsalibrate}.

\section{Results}
\label{sec:results}

\begin{figure}
\begin{center}
  \includegraphics[width=80mm]{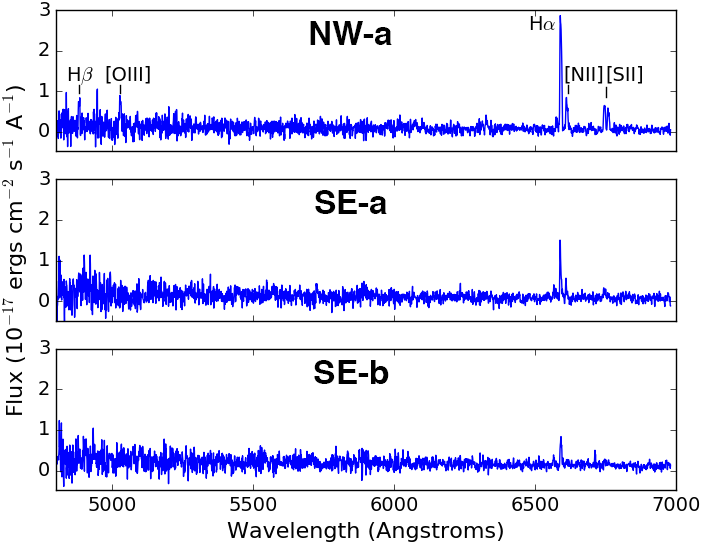}
  \caption{GMOS spectra of three regions in AGC~208457.  Prominent emission lines are labelled in the top panel.
\label{fig:spectra1}}
\end{center}
\end{figure}

\begin{figure}
\begin{center}
  \includegraphics[width=80mm]{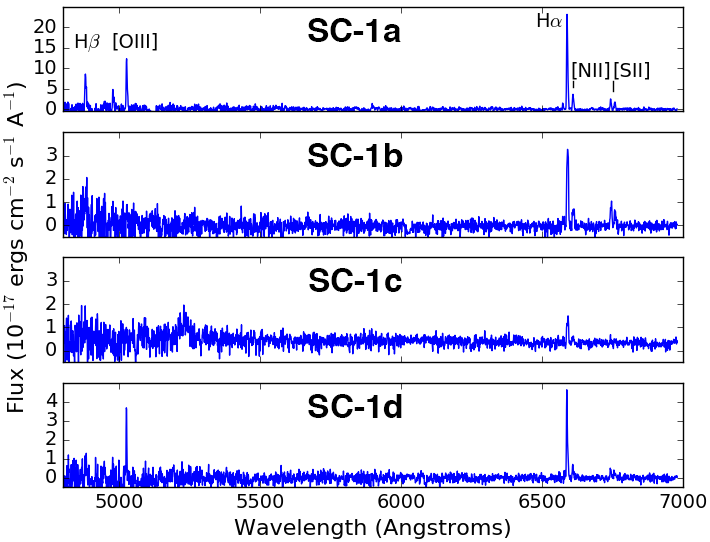}
  \caption{GMOS spectra of four regions in SC-1 in the NGC~4747 tail.  Prominent emission lines are labelled in the top panel.
\label{fig:spectra2}}
\end{center}
\end{figure}

In addition to the properties previously measured from the H{\sc{i}} and optical photometry observations of AGC~208457 and the SCs in the gas-rich tail of NGC~4747, optical velocities as well as gas-phase metallicity measurements are required in order to confirm tidal origins.  Figures \ref{fig:spectra1} and \ref{fig:spectra2} show the spectra that were extracted from the brightest optical regions along each slit in our observations.  For AGC~208457, there is one spectrum for the northwest (NW) and two for the southeast (SE) stellar clumps, as indicated by the arrows in Figure~\ref{fig:TDG}.  We were able to extract four spectra from SC-1 in the NGC~4747 tail (see Figure~\ref{fig:knots}); however, SC-2 was too faint to detect emission lines. As evident in the spectra, there appears to be minimal (if any) continuum emission associated with the sources.

The H$\alpha$, H$\beta$, [{N\sc{ii}}], [{O\sc{iii}}] and [{S\sc{ii}}] emission line fluxes were measured from each spectrum using the {\sc splot} routine within IRAF and are presented in Table~\ref{table:flux}.  The dominant source of uncertainty arises from the sensitivity function that is derived from standard stellar spectra.  The residuals in fitting the sensitivity function translate to a $\sim$10-15\% uncertainty in the measured line fluxes.  Note, the spectrum for SC-1c shows a broad emission feature at $\sim$5250\AA, which is likely a background source contaminating the noisier region of the data.

\begin{table*}
 \centering
 \begin{minipage}{120mm}
 \caption{Emission line fluxes for various regions in AGC 208457 and SC-1 in the NGC~4747 tail.}
 \label{table:flux}
\begin{tabular}{ l c c c c c c c c}   
\hline									
						&\multicolumn{8}{c}{Flux ($\times 10^{-17}$ erg cm$^{-2}$ s$^{-1}$)}\\
\hline
						&\multicolumn{3}{c}{AGC 208457} &| &\multicolumn{4}{c}{NGC 4747 tail}\\
						& NW-a	& SE-a  	& SE-b 	&|& SC-1a& SC-1b	& SC-1c	& SC-1d\\
\hline
H$\beta$					&	3.4	$\pm$	0.5	&	1.1	$\pm$	0.2	&	--				&	|	&	48	$\pm$	7	&	8.0 	$\pm$	0.8	&	--		&	6.0	$\pm$	0.9	\\
{[{O\sc{iii}}]$_{\lambda 4959}$}	&	1.4	$\pm$	0.2	&	--				&	--				&	|	&	19	$\pm$	3	&	--				&	--		&	--			\\
{[{O\sc{iii}}]$_{\lambda 5007}$}	&	5.6	$\pm$	0.8	&	--				&	--				&	|	&	59	$\pm$	9	&	4.0	$\pm$ 	0.6	&	--		&	12	$\pm$	2	\\
{[{N\sc{ii}}]$_{\lambda 6548}$}	&	2.2	$\pm$	0.3	&	0.5	$\pm$	0.1	&	--				&	|	&	6.0	$\pm$	0.9	&	--				&	--		&	--			\\
H$\alpha$					&	21	$\pm$	2	&	6.8	$\pm$	0.7	&	4.3	$\pm$	0.4	&	|	&	120	$\pm$	10	&	25	$\pm$	3	&	8.0 $\pm$ 0.8	&	20	$\pm$	2	\\
{[{N\sc{ii}}]$_{\lambda 6583}$}	&	6.1	$\pm$	0.6	&	1.7	$\pm$	0.2	&	0.5	$\pm$	0.1	&	|	&	20	$\pm$	2	&	6.0	$\pm$	0.6	&	2.0 $\pm$ 0.3	&	3.0	$\pm$	0.5	\\
{[{S\sc{ii}}]$_{\lambda 6717}$}	&	4.8	$\pm$	0.5	&	0.5	$\pm$	0.1	&	0.5	$\pm$	0.1	&	|	&	14	$\pm$	1	&	6.0	$\pm$	0.9	&	1.0 $\pm$ 0.2	&	2.0	$\pm$	0.3	\\
{[{S\sc{ii}}]$_{\lambda 6731}$}	&	3.6	$\pm$	0.4	&	0.5	$\pm$	0.1	&	0.3	$\pm$	0.1	&	|	&	12	$\pm$	1	&	4.0 	$\pm$ 	0.6	&	--		&	--			\\
\hline																
\end{tabular} 
\end{minipage}
\end{table*}

The electron temperature, directly calculated from the flux of [{O\sc{iii}}]$_{\lambda 4363}$, is considered one of the most straight-forward methods to determine the gas-phase metallicity of a source \citep{ly2014}.  Although the spectral range of our observations included [{O\sc{iii}}]$_{\lambda 4363}$, this line is too weak to detect in our sources, which is typical of dwarf galaxies (see \citealt{cro2009}).  As such, we follow the method employed by \citet{duc2014} and use the empirical N2 and O3N2 line ratios defined as
\begin{equation}
\mbox{N2} = \log \left( \frac{[{\mbox{N\sc{ii}}}]_{\lambda 6583}}{\mbox{H}\alpha} \right)
\end{equation}
\begin{equation}
\mbox{O3N2} = \log \left(\frac{[{\mbox{O\sc{iii}}}]_{\lambda 5007}}{\mbox{H}\beta} \times \frac{\mbox{H}\alpha}{[{\mbox{N\sc{ii}}}]_{\lambda 6583}} \right).
\end{equation}
We assume the solar oxygen abundance to be $12 + \log$(O/H)$_{\odot}$ = 8.65 \citep{pil2007} and use the calibrations from \citet{mar2013} to estimate the chemical abundances from the line ratios:
\begin{equation}
 \label{eq:N2}
12 + \log\mbox{(O/H)} = 8.743 + 0.462 \times \mbox{N2}
\end{equation}
\begin{equation}
 \label{eq:O3N2}
12 + \log\mbox{(O/H)} = 8.533 - 0.214 \times \mbox{O3N2}.
\end{equation}

The properties for AGC~208457 and the gas-rich tail of NGC~4747 are presented in Table \ref{table:properties}.  Radial {H\sc{i}} velocities (v$_{H_I}$) and velocity widths (W$_{20}$) were measured from the GMRT data (\citetalias{lee2012} and \citetalias{lee2016}).  Optical velocities (v$_{\mbox{opt}}$) are the average of the radial velocity values for all detectable emission lines along the optical spectra extracted for each source from the GMOS-N observations.  The uncertainty in v$_{\mbox{opt}}$ reflects the standard deviation of the averaged velocity values and the resolution of the spectra.  Due to the faintness of the [{O\sc{iii}}]$_{\lambda 5007}$ emission line, O3N2 ratios could not be computed for all optical sources.  Assuming the group distances for each source, their star-formation rates (SFRs) were estimated using:
\begin{equation}
\mbox{SFR$_{\mbox{H}\alpha}$[$M_{\odot}$ yr$^{-1}$]} = 7.9 \times 10^{-42} L_{\mbox{H}\alpha}
\end{equation}
where $L_{\mbox{H}\alpha}$ is the luminosity in erg s$^{-1}$ \citep{ken1998}.

We attempted to use the spectral synthesis code {\sc{Cloudy}} v.C17 \citep{fer2017} to reproduce the spectra for each source.  However, due to the limited number of detected emission lines, the software was unable to converge on physically possible solutions.  Additionally, since all the sources are quite faint and lack strong continuum emission, deeper analysis of any older stellar populations within these systems is not feasible with the current data (see \citealt{ber2012}, \citealt{sco2018}).

\begin{table*}
 \centering
 \begin{minipage}{180mm}
 \caption{Properties of AGC 208457 and the features in the NGC~4747 tail.  Values indicated with $^{\dagger}$ are from \citetalias{lee2012} and \citetalias{lee2016}}
 \label{table:properties}
\begin{tabular}{ l @{} c @{} c c c  c  c c @{} c c c c}   
\hline
	&	\multicolumn{4}{c}{AGC 208457} &| &\multicolumn{6}{c}{NGC 4747 tail}\\									
	&	{H\sc{i}} cloud		& NW-a		& SE-a  		& SE-b 		&|&	NE-1	&	NE-2	&	SC-1a	&	SC-1b	&	SC-1c	&	SC-1d\\
\hline									
v$_{H_I}$ (km s$^{-1}$)$^{\dagger}$	&1343 $\pm$ 3	&	--	&	--	&--	&|&	1265 $\pm$ 3	&	1258 $\pm$ 3	&	--	&	--	&	--	&	--	\\
W$_{20}$ (km s$^{-1}$)$^{\dagger}$	&33 $\pm$ 7	&	--	&	--	&--	&|&	49 $\pm$ 5	&	48 $\pm$ 5	&	--	&	--	&	--	&	--	\\
v$_{\mbox{opt}}$ (km s$^{-1}$)	&--	&1300 $\pm$ 50	&1220 $\pm$ 50	&1330	 $\pm$ 60		&|&	--	&--	&1180 $\pm$ 60	&1240 $\pm$ 60	&1220 $\pm$ 80	&1170 $\pm$ 70\\
12+log(O/H) - O3N2	&	--	&	8.4 $\pm$ 0.2	&	--	&	--		&|&	--	&	--	&	8.4 $\pm$ 0.2	&	8.5 $\pm$ 0.2	&	--	&	8.3 $\pm$ 0.2	\\
12+log(O/H) - N2	&	--	&	8.5 $\pm$ 0.2	&	8.5 $\pm$ 0.2	&	8.3 $\pm$ 0.3	&|&	--	&	--	&	8.4 $\pm$ 0.2	&	8.5 $\pm$ 0.2	&	8.5 $\pm$ 0.3	&	8.4 $\pm$ 0.2	\\
SFR$_{\mbox{H}\alpha}$ ($M_{\odot}$ yr$^{-1}$)
&	--	& $1\times10^{-4}$	& $3\times10^{-5}$	& $2\times10^{-5}$	&|&	--	&	--& $2\times10^{-4}$	& $4\times10^{-5}$	& $1\times10^{-5}$	& $3\times10^{-5}$\\
\hline																	
\end{tabular} 
\end{minipage}
\end{table*}

\begin{figure}
\begin{center}
  \includegraphics[width=80mm]{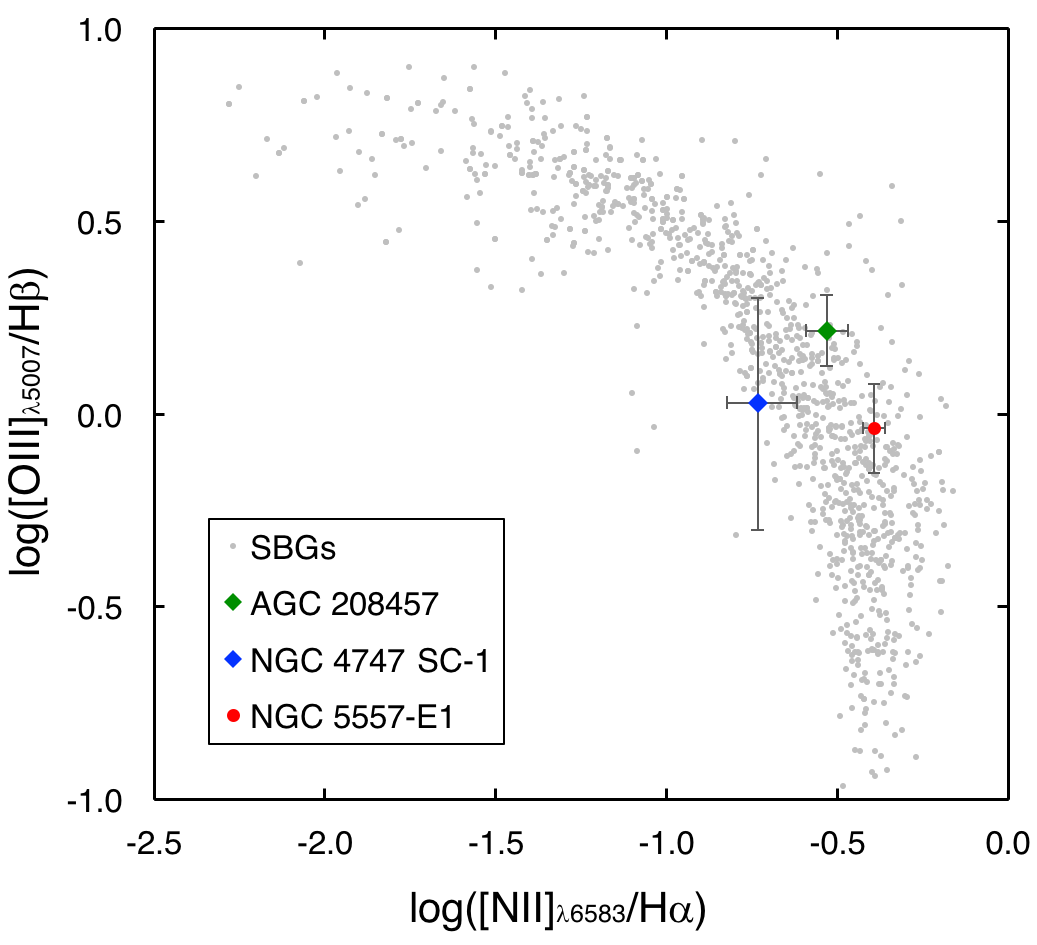}
  \caption{Spectral excitation diagnostic diagram showing the line ratio of AGC 208457 and the average value for SC-1 in the NGC~4747 tail in comparison with SBGs from the KISS sample (\citealt{weg2003}, \citealt{gro2004}, \citealt{jan2005}, \citealt{sal2005}) and confirmed TDG, NGC~5557-E1 \citep{duc2014}.
\label{fig:sbg} }
\end{center}
\end{figure}

Figure~\ref{fig:sbg} plots the available line ratio values of AGC 208457 and SC-1 in the NGC~4747 tail on a spectral excitation diagnostic diagram.  The data point for SC-1 is an average of the measured values from the various regions of the stellar cluster and the error bars indicate the range of these values.  The grey data points are from the Kitt Peak National Observatory (KPNO) International Spectroscopic Survey (KISS) for starbursting galaxies (SBGs; \citealt{weg2003}, \citealt{gro2004}, \citealt{jan2005}, \citealt{sal2005}).  High metallicity galaxies are found in the lower right while low metallicity galaxies are in the upper left of the plot.  NGC~5557-E1, a dwarf galaxy that was confirmed to be tidal with similar optical spectroscopy observations as employed in this paper (see \citealt{duc2014}), has been included for comparison.

\section{Discussion}
\label{sec:discuss}

The confirmation of a tidally-formed dwarf galaxy requires a suite of observations to verify longevity and pre-enriched content.  In \citetalias{lee2012} and \citetalias{lee2016}, we identified three {H\sc{i}} objects, AGC~208457 around the NGC 3166/9 pair and two gas-rich knots NE-1 and NE-2 in the tail of NGC~4747, each with sufficient gas mass for long-term survivability and that appear to be in the vicinity of fairly young stellar features.  In this paper, we presented optical spectroscopy of these features that confirm enriched metallicity and verify a physical association between the stars and gas, assuming similar radial velocities of the {H\sc{i}} and stellar components (which already spatially coincide) imply that the line emissions stem from a common astronomical source.  

\subsection{AGC~208457, a TDG candidate}

The optical velocities measured for the three components in AGC~208457 are 10-120 km s$^{-1}$ lower than the GMRT {H\sc{i}} velocity of the cloud (v$_{H_I} =1343 \pm 3$ km s$^{-1}$).  Stellar clumps NW-a and SE-b (v$_{\mbox{opt}} = 1300 \pm 50$ km s$^{-1}$ and $1330 \pm 60$ km s$^{-1}$) agree within the uncertainties and velocity range of the {H\sc{i}} component; whereas, SE-a has a significantly lower value (v$_{\mbox{opt}} = 1220 \pm 50$ km s$^{-1}$).  SE-a is possibly stellar debris that is outside of the TDG candidate or indicates an unvirialized system.  Nevertheless, this source is still within the measured  {H\sc{i}} velocity range for NGC~3166/9 pair (see \citetalias{lee2012}).

The highest oxygen abundances ($8.5 \pm 0.2$) are found in NW-a and SE-a, while SE-b ($8.3 \pm 0.2$) is lower. These metallicity values are comparable to other tidal features found the literature (i.e. 12+log(O/H)~>~8.3; \citealt{duc1998}, \citealt{dem2012}, \citealt{lel2015}).  Additionally, on the spectral excitation diagram (Fig.~\ref{fig:sbg}), the NW stellar clump of AGC~208457 has line ratios that are consistent with metal-rich SBGs, which are located in the bottom right portion of the plot.  While the metallicity values alone do not unambiguously distinguish AGC~208457 as a TDG, this object is relatively metal-rich when compared to typical dwarf irregular galaxies (see \citealt{van2006}, \citealt{cro2009}, \citealt{ber2012}).

The enhanced metallicity of AGC~208457 is likely inherited from its parent galaxies. Previous spectroscopy observations by \citet{sil2006} of NGC~3166 and NGC~3169 indicate that their central regions are metal-rich with above-solar metallicity.  However, the mean metallicity of the outer disk region of NGC~3166 is fairly low \citep{sil2012}.  NGC~3166 has a fairly undisturbed stellar disk; whereas, both the {H\sc{i}} and stellar components of NGC~3169 show clear signs of the gravitational interaction occurring between the two galaxies \citepalias{lee2016}.  Based on the {H\sc{i}} scaling relations from \citet{den2014}, NGC~3169 has a standard gas mass for a galaxy of its size and colour but NGC~3166 has no discernible {H\sc{i}} counterpart and is therefore fairly {H\sc{i}}-deficient \citepalias{lee2012}.  It appears that AGC~208457 has been formed from material likely drawn out of NGC~3166 and NGC~3169.

The H$\alpha$ SFR estimated for AGC~208457 is 1-2 orders of magnitude lower than the UV SFR (estimated in \citetalias{lee2016}), which is typical for gas-rich dwarf galaxies (see \citealt{lee2009}).  Overall, with an {H\sc{i}} mass of M$_{H_I} = 2.3 \pm 0.3 \times 10^8$ $M_{\odot}$, a negligible dark matter component (assuming that the object is self-gravitating; \citetalias{lee2012}), fairly young stars (\citetalias{lee2016}), a high metallicity, and a plausible formation scenario, AGC~208457 has all the properties of a bonafide TDG.

\subsection{Potential TDGs in the gas-rich tail of NGC~4747}

The gaseous tail emerging from NGC~4747 is modestly offset from a bent stellar tail located in the same spatial region \citepalias{lee2016}.  The optical velocities measured for the four stellar components within the tail of NGC~4747 are 20-100 km s$^{-1}$ lower than the GMRT {H\sc{i}} velocity of the two {H\sc{i}} peaks (v$_{H_I} =1265 \pm 3$ km s$^{-1}$ and $1258 \pm 3$ km s$^{-1}$).  Nevertheless, the values agree within the uncertainties and {H\sc{i}} velocity range, signifying an actual physical association between the stellar and {H\sc{i}} tails.  SC-1a and SC-1d do have lower central velocities, which may indicate that their stars are projected in front of the {H\sc{i}} tail.  Possible projection effects from the bent tail precludes clear source separation and complicates deeper analysis of the contents of this tidal feature.  Although we could only extract spectra from the region originally identified as SC-1, we presume that these stars are located within the stellar tail of NGC~4747 and their properties are representative of the entire tail.  

The highest oxygen abundances ($8.5 \pm 0.3$) are found in SC-1b and SC-1c, while the other two regions are around $8.4 \pm 0.2$. We note that both methods (O3N2 and N2) produce consistent results.  As expected, the stars in the NGC~4747 tail are fairly young (< 2.5 Gyr; see \citetalias{lee2016}) and metal-rich, as shown in Fig.~\ref{fig:sbg}, thereby substantiating tidal origins.  This system is in a much earlier evolutionary stage than AGC~208457 and turbulent motions, which can inhibit star-formation in high density {H\sc{i}} regions (see \citealt{sen2015}), are still present within the tail.  Nevertheless, NE-1 and NE-2 appear to have sufficient mass -- especially if they continue to accrete content from the tidal tail (see \citetalias{lee2016}) -- to eventually decouple from their host and become self-gravitating TDGs.

\section{Conclusions}
\label{sec:conclude}

Using GMOS long-slit spectroscopy, we confirm the stellar counterparts and tidal origins of two {H\sc{i}} sources, AGC 208457 and the gas-rich tail of NGC~4747, located around the outskirts of two interacting galaxy pairs (NGC~3166/9 and NGC~4725/47 respectively).  The optical velocities of the stellar components are consistent with the {H\sc{i}} velocities, signifying physical associations between the features.  We measure high oxygen abundances in these objects, indicating that they have formed out of pre-enriched material drawn from their parent galaxies during an interaction event.  In conjunction with the previously measured {H\sc{i}}, dynamical, and stellar properties, AGC~208457 has the characteristics of a genuine TDG.  The {H\sc{i}} knots embedded in the tidal tail of NGC~4747 are in an earlier stage of evolution and they have the potential to evolve into long-lived TDGs.

\section*{Acknowledgements}

We thank the reviewer for his/her thoroughly detailed comments and suggestions to improve the clarity of this paper.  The data presented in this paper are based on observations obtained (during program GN-2016B-Q-24) at the Gemini Observatory, which is operated by the Association of Universities for Research in Astronomy, Inc., under a cooperative agreement with the NSF on behalf of the Gemini partnership: the National Science Foundation (United States), the National Research Council (Canada), CONICYT (Chile), Ministerio de Ciencia, Tecnolog\'{i}a e Innovaci\'{o}n Productiva (Argentina), and Minist\'{e}rio da Ci\^{e}ncia, Tecnologia e Inova\c{c}\~{a}o (Brazil).  The authors wish to recognize and acknowledge the very significant cultural role and reverence that the summit of Mauna Kea has always had within the indigenous Hawaiian community.  We are most fortunate to have the opportunity to obtain observations from this mountain.





\begin{thebibliography}{99}
\bibitem[\protect\citeauthoryear{Berg et al.}{2012}]{ber2012}
Berg D.A., et al., 2012, ApJ, 754, 99 
\bibitem[\protect\citeauthoryear{Blanton \& Moustakas}{2009}]{bla2009} 
Blanton M.R., Moustakas J., 2009, ARA\&A, 47, 159
\bibitem[\protect\citeauthoryear{Bournaud \& Duc}{2006}]{bou2006} 
Bournaud F., Duc P.-A., 2006, A\&A, 456, 481
\bibitem[\protect\citeauthoryear{Bournaud et al.}{2007}]{bou2007}
Bournaud F., et al., 2007, Science, 316, 1166 
\bibitem[\protect\citeauthoryear{Croxall et al.}{2009}]{cro2009}
Croxall K.V., van Zee L., Lee H., Skillman E.D., Lee J.C., Cote S., Kennicutt R.C. Jr., Miller B.W., 2009, ApJ, 705, 723
\bibitem[\protect\citeauthoryear{de Mello et al.}{2012}]{dem2012}
de Mello D.F., Urrutia-Viscarra F., Mendes de Oliveira C., Torres-Flores S., Carrasco E. R., Cypriano E., 2012, MNRAS, 426, 2441
\bibitem[\protect\citeauthoryear{Denes, Kilborn \& Koribalski}{2014}]{den2014}
Denes H., Kilborn V.A., Koribalski B.S., 2014, MNRAS, 444, 667
\bibitem[\protect\citeauthoryear{Duc \& Mirabel}{1998}]{duc1998} 
Duc P.-A., Mirabel I.F., 1998, A\&A, 333, 813
\bibitem[\protect\citeauthoryear{Duc et al.}{2014}]{duc2014} 
Duc P.-A., Paudel S., McDermid R. M., Cuillandre J.-C., Serra P., Bournaud F., Cappellari M., Emsellem, E., 2014, MNRAS, 440, 1458	
\bibitem[\protect\citeauthoryear{Ferland et al.}{2017}]{fer2017}
Ferland G.J., et al., 2017, RMxAA, 53, 385 
\bibitem[\protect\citeauthoryear{Flores et al.}{2016}]{flo2016}
Flores H., Hammer F., Fouquet S., Puech M., Kroupa P., Yang Y., Pawlowski M., 2016, MNRAS, 457, 14
\bibitem[\protect\citeauthoryear{Gronwell et al.}{2004}]{gro2004}
Gronwall C., Jangren A., Salzer J., Werk J., Ciardullo R., 2004, AJ, 128, 644
\bibitem[\protect\citeauthoryear{Jangren et al.}{2005}]{jan2005}
Jangren A., Wegner G., Salzer J., Werk J., Gronwall C., 2005, AJ, 130, 496
\bibitem[\protect\citeauthoryear{Kennicutt}{1998}]{ken1998}
Kennicutt Jr., R. C., 1998, ARA\&A, 36, 189
\bibitem[\protect\citeauthoryear{Kroupa}{2012}]{kro2012} 
Kroupa P., 2012, PASA, 29, 395
\bibitem[\protect\citeauthoryear{Lee et al.}{2009}]{lee2009}
Lee J.C., et al., 2009, ApJ, 706, 599 
\bibitem[\protect\citeauthoryear{Lee-Waddell et al.}{2012}]{lee2012}
Lee-Waddell K., Spekkens K., Haynes M.P., Stierwalt S., Chengalur J., Chandra P., Giovanelli R., 2012, MNRAS, 427, 2314 (Paper 1)
\bibitem[\protect\citeauthoryear{Lee-Waddell et al.}{2016}]{lee2016}
Lee-Waddell K., et al., 2016, MNRAS, 460, 2945 (Paper 2)
\bibitem[\protect\citeauthoryear{Lelli et al.}{2015}]{lel2015}
Lelli F., et al., 2015, A\&A, 584, 113
\bibitem[\protect\citeauthoryear{Lisenfeld et al.}{2016}]{lis2016}
Lisenfeld U., Braine J., Duc P.-A., Boquien M., Brinks E., Bournaud F., Lelli F., Charmandaris V., 2016, A\&A, 590, 92
\bibitem[\protect\citeauthoryear{Ly et al.}{2014}]{ly2014}
Ly C., Malkan M., Nagao T., Kashikawa N., Shimasaku K., Hayashi M., 2014, ApJ, 780, 122
\bibitem[\protect\citeauthoryear{Madrid \& Donzelli}{2013}]{mad2013}
Madrid J.P., Donzelli C.J., 2013, ApJ, 770, 158
\bibitem[\protect\citeauthoryear{Marino et al.}{2013}]{mar2013}
Marino R.A., et al., 2013, A\&A, 559, 114
\bibitem[\protect\citeauthoryear{Maybhate et al.}{2007}]{may2007}
Maybhate A., Masiero J., Hibbard J., Charlton J., Palma C., Knierman K., English J., 2007, MNRAS, 381, 59
\bibitem[\protect\citeauthoryear{Pilyugin, Thuan \& Vilchez}{2007}]{pil2007}
Pilyugin L.S., Thuan T.X., Vilchez J.M., 2007, MNRAS, 376, 353
\bibitem[\protect\citeauthoryear{Salzer et al.}{2005}]{sal2005}
Salzer J., Jangren A., Gronwall C., Werk J., Chomiuk L., Caperton K., Melbourne J., McKinstry K., 2005, AJ, 130, 2584
\bibitem[\protect\citeauthoryear{Scott et al.}{2018}]{sco2018}
Scott T.C., et al., 2018, MNRAS, 475, 1148
\bibitem[\protect\citeauthoryear{Sengupta et al.}{2015}]{sen2015}
Sengupta C., Scott T.C., Paudel S., Saikia D.J., Dwarakanath K.S., Sohn B.W., 2015, A\&A, 584, 114
\bibitem[\protect\citeauthoryear{Sil'chenko \& Afanasiev}{2006}]{sil2006}
Sil'chenko O.K., Afanasiev V.L., 2006, Astron. Lett., 32, 534
\bibitem[\protect\citeauthoryear{Sil'chenko et al.}{2012}]{sil2012}
Sil'chenko O.K., Proshina I.S., Shulga A.P., Koposov S.E., 2012, MNRAS, 427, 790
\bibitem[\protect\citeauthoryear{Tully et al.}{2013}]{tul2013}
Tully B., et al., 2013, AJ, 146, 86 
\bibitem[\protect\citeauthoryear{Toomre \& Toomre}{1972}]{too1972}
Toomre A., Toomre J., 1972, ApJ, 178, 623
\bibitem[\protect\citeauthoryear{van Zee \& Haynes}{2006}]{van2006}
van Zee L., Haynes M.P., 2006, ApJ, 636, 214
\bibitem[\protect\citeauthoryear{Wegner et al.}{2003}]{weg2003}
Wegner G., Salzer J., Jangren A., Gronwall C., Melbourne J., 2003, AJ, 125, 2373
\end{thebibliography}



\bsp	
\label{lastpage}
\end{document}